# Quasi-Phase-Matched Second Harmonic Generation in Poling-Free Lithium Niobate Microrings with Surface Molecular Pattern

*Jinsong Cheng*[1,#], *Simin Yu*[2,#], *Jiahe Wang*[1], *Yilun Zhang*[1], *Juanjuan Lu*[2,*], *Xiaoqin Shen*[1,*]

[1]State Key Laboratory of Quantum Functional Materials, School of Physical Science and Technology, ShanghaiTech University, China 201210

[2]School of Informational Science and Technology, ShanghaiTech University, China 201210

# The authors contributed equally to this work.

Corresponding email: lujj2@shanghaitech.edu.cn; shenxq@shanghaitech.edu.cn

**Abstract**

Thin film lithium niobate (TFLN) is a promising integrated photonic platform owing to its large quadratic nonlinearity enabling vital applications such as second-harmonic generation (SHG), optical parametric amplification, and spontaneous parametric down-conversion. However, to achieve high SH conversion efficiency, quasi-phase matching (QPM) through periodic electrical poling is typically required, which remains challenging in certain applications like multiple-ring coupled devices. Here, we demonstrate a surface molecule patterning strategy to enable QPM in a poling-free TFLN microring for SHG. An on-chip normalized conversion efficiency of approximately 11.5% $W^{-1}$ is obtained in the poling-free TFLN microring with surface molecular pattern, showing an approximate 400-fold improvement compared to the same microring without surface molecular pattern. This approach offers a versatile route to engineer interfacial nonlinear interactions for nonlinear optics and quantum photonics applications and may be extended to multiple-ring coupled resonators and other platforms.

Keywords: quasi-phase matching, periodic surface molecular pattern, lithium niobate, second-harmonic generation, microring resonator, integrated nonlinear photonics

## 1 | Introduction

Second-order nonlinearity $\chi^{(2)}$, is a cornerstone for many optical processes, including second-harmonic generation (SHG),[1-3] sum- and difference- frequency generation,[4] spontaneous parametric down-conversion.[5] Among these, SHG is of particularly important for a wide range of classical and quantum optical applications such as quantum light generation,[6] optical atomic clocks,[7] frequency self-referencing,[8] optical-frequency synthesis.[9] With the advancement of micro-nano fabrication technology, a variety of integrated photonics platforms have emerged,[10] such as silicon nitride,[11-13] aluminium nitride,[14, 15] gallium arsenide,[16, 17] silicon carbide,[18, 19] thin film lithium niobate (TFLN).[20] Among the various nonlinear materials, TFLN is a highly attractive platform for integrated photonics due to its strong optical field confinement, broad transparency window (0.35–5.0 μm), and flexible ferroelectric domain engineering.[21, 22] More importantly, TFLN exhibits a large second-order nonlinear susceptibility and a strong electro-optic effect, enabling efficient frequency conversion as well as high-speed, low-voltage on-chip modulation.[23-26] These advantages make TFLN a versatile and scalable foundation for compact light sources, modulators, and nonlinear photonic circuits.[27-29]

The phase-matching condition is essential for realizing efficient quadratic nonlinear processes.[30] In recent decades, cavity-enhanced SHG based on modal phase matching (MPM)[31-33] and quasi-phase matching (QPM)[34-36] have been extensively studied. Compared with MPM, QPM places far fewer constraints on the material refractive index and waveguide geometry. Phase matching across different wavelength bands and poling configurations can be achieved simply by selecting an appropriate domain-inversion period. This method enables independent optimization of key factors such as the effective nonlinear coefficient, modal overlap, and device dimensions, making QPM a widely adopted approach in practice to balance high efficiency with design versatility. However, large-scale, high-yield fabrication of on-chip periodically poled lithium niobate (PPLN) devices remains complicated, particularly for multiple-ring coupled TFLN devices that are promising for quantum nonlinear photonics.[2, 37] Therefore, it is of great interest to develop new approaches to realize QPM in poling-free TFLN microrings for SHG.

Here, we demonstrate a surface molecule patterning strategy to enable QPM in a poling-free TFLN microring for SHG. The surface molecular pattern modulates an additional interfacial nonlinear response in the poling-free TFLN microrings, resulting in a nonlinear grating effect to compensate the azimuthal mode-number mismatch. Experimentally, a poling-free z-cut TFLN microring was functionalized with periodical surface pattern of nonlinear optical (NLO) molecules, and was coupled to a single pulley waveguide for dual-band operation. The detuning between pump and second-harmonic mode is compensated by on-chip thermal control to realize double-resonance in the microrings. The poling-free TFLN microring effectively generates SHG emission, with on-chip second-harmonic normalized conversion efficiency of about 11.5% $W^{-1}$, showing about 400-fold improvement compared to the same TFLN microring without surface molecular pattern. Optimization of the surface molecular pattern would further improve the conversion efficiency. This approach offers a versatile route to engineer interfacial nonlinear interactions for nonlinear optics and quantum photonics applications and may be extended to multiple-ring coupled resonators and other platforms.

## 2 | Theoretical analysis and design principle

We consider a $z$-cut LN microring supporting the fundamental and SH $TM_{00}$ modes with resonance frequencies $\omega_a$ and $\omega_b$, respectively. For a pump at $\omega_p$, the on-chip normalized SHG conversion efficiency in the undepleted pump regime is given by[38]

$$\eta = \frac{P_{SHG}}{P_p^2} = \frac{|g|^2\omega_b}{\hbar\omega_p^2}\frac{2\kappa_{c,b}}{\delta_b^2+\kappa_b^2}\left(\frac{2\kappa_{c,a}}{\delta_a^2+\kappa_a^2}\right)^2, \quad (1)$$

where $P_p$ and $P_{SHG}$ denote the on-chip pump power and SH power extracted into the pulley waveguide, respectively. The external coupling, intrinsic loss, and total decay rate of each cavity mode are denoted by $\kappa_c$, $\kappa_0$ and $\kappa = \kappa_c + \kappa_0$, respectively. The detunings are defined as $\delta_a = \omega_a - \omega_p$ and $\delta_b = \omega_b - 2\omega_p$. The nonlinear coupling strength $g$ between mode a and b given by[39]

$$g \approx \frac{3}{4\sqrt{2}}\sqrt{\frac{\hbar\omega_a^2\omega_b}{\varepsilon_0\,2\pi R}}\frac{\zeta}{\varepsilon_a\sqrt{\varepsilon_b}}\chi_{eff}^{(2)}\delta(m_b - 2m_a - M), \quad (2)$$

$R$ is the microring radius, $\varepsilon_0$ is the vacuum permittivity, $\zeta$ is the transverse nonlinear overlap factor, $\varepsilon_{a(b)}$ and $m_{a(b)}$ are the effective relative permittivity and azimuthal number of mode $a(b)$ in the microring, and the integer $M = \frac{2\pi R}{\Lambda}$ is defined by the surface molecular pattern, with $\Lambda$ being the pattern period.

The QPM condition $m_b - 2m_a - M = 0$ can be satisfied by surface functionalization of a periodical NLO molecule pattern, which introduces an additional interfacial nonlinear response in the covered regions while leaving the underlying TFLN susceptibility unchanged. Within the adopted effective-response model,[40] the ratio of the surface evanescent field $\rho$, the susceptibility of surface molecules $\chi_m^{(2)}$ and the patterning duty circle D determine the effective $\chi_{eff}^{(2)}$ of the surface molecule patterned poling-free TFLN with the N[th] order QPM, through the relation

$$\chi_{eff}^{(2)} = 2\rho^2 \chi_m^{(2)} \frac{sin(\pi N D)}{N\pi} \tag{3}$$

where $\chi_m^{(2)} \propto N_s \beta_{uuu} \langle cos^3\theta \rangle$ is the susceptibility of surface NLO molecules, which is proportional to the local molecular surface density $N_s$, the dominant molecular hyperpolarizability $\beta_{uuu}$, and the molecular tilt angle relative to the surface normal $\theta$ within the covered regions. Thus, for the first order of QPM, the coupling strength is given by

$$g_{QPM} \approx \frac{3}{4\sqrt{2}} \sqrt{\frac{\hbar\omega_a^2\omega_b}{\varepsilon_0\, 2\pi R}} \frac{\zeta}{\varepsilon_a\sqrt{\varepsilon_b}} \rho^2 \chi_m^{(2)} \frac{sin(\pi D)}{\pi}, \tag{4}$$

which indicates QPM SHG process can be enabled by surface patterned NLO molecules on a ring resonator. This mechanism modulates the surface NLO molecular response between covered and uncovered regions rather than ferroelectric domains of TFLN, offering a surface molecular pattern enabled QPM for SHG in poling-free TFLN resonators.

Figure 1 depicts the design principle of the pulley-coupled, poling-free TFLN microring resonator with periodical surface NLO molecular pattern, where the SHG process involves the fundamental transverse magnetic modes $TM_{00}$ of angular frequency $\omega_a$ and $\omega_b$ at 1550 nm and 775 nm, respectively. The surface of the poling-free TFLN microring is functionalized with a periodical pattern of NLO molecules, 4-[4-diethylamino(styryl)] pyridinium (DASP) which interact with the pump via the optical evanescent field.

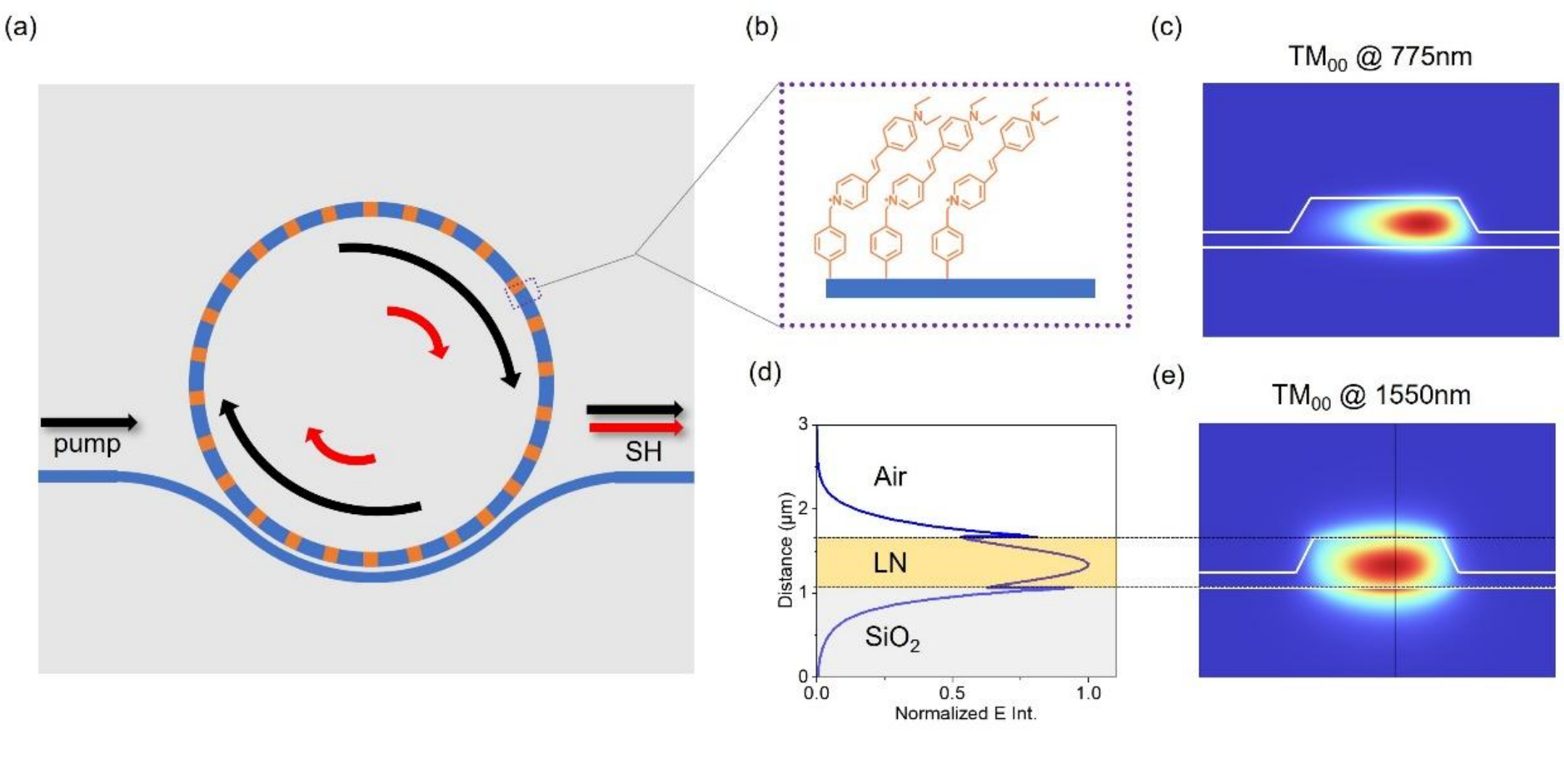

FIGURE 1 | (a-b) Schematic of the z-cut poling-free TFLN microring resonator with periodical molecular functionlization for SHG and its zoomed view, orange regions

indicate selectively grafted NLO molecules on the surface. (c) Simulated mode profiles of the SH $TM_{00}$ mode. (d) The mode profile at the device cross section indicated by the solid black line in (e), confirms that the optical field extends into the surrounding environment, enabling interaction with the DASP monomolecular layer on the surface. (e) Simulated mode profiles of the $TM_{00}$ mode.

## 3 | Results and Discussion

Figure 2a shows the geometries of the poling-free TFLN microring resonator with the radius of 70 μm and the top width of 1.8 μm. The width of the pulley waveguide is 600 nm, and the coupling gap is 400 nm. Cross-sectional SEM images show a trapezoidal waveguide profile with a sidewall angle of approximately 60°. Figure 2b shows the simulated dispersion curves of the effective refractive indices for the $TM_{00}$ mode at the FW and SH. A first-order QPM scheme using TM-polarized modes was implemented, which requires $\Lambda = \lambda_a/2(n_b - n_a)$. The optimal molecular grafting period was calculated to be Λ = 2.94 μm, corresponding to an integer $M$ is 150. Because their effective refractive indices differ significantly, natural phase matching cannot satisfy the momentum conservation condition for SHG.

The surfaces of the TFLN microring resonators were functionalized with DASP using a two-step molecular self-assembly method modified from Ref. [41], as illustrated in Figure 2c. From this lithographically defined pattern, we extracted a molecular period of Λ = 2.94 μm with a duty cycle of approximately 35%. Detailed procedures for the functionalization process are provided in the Supplementary Information. The high-resolution X-ray photoelectron spectroscopy spectrum of the DASP-grafted LN clearly shows a binding energy peak at 399.4 eV from nitrogen atoms (Figure 2d), confirming the formation of the DASP molecular layer on the surface.

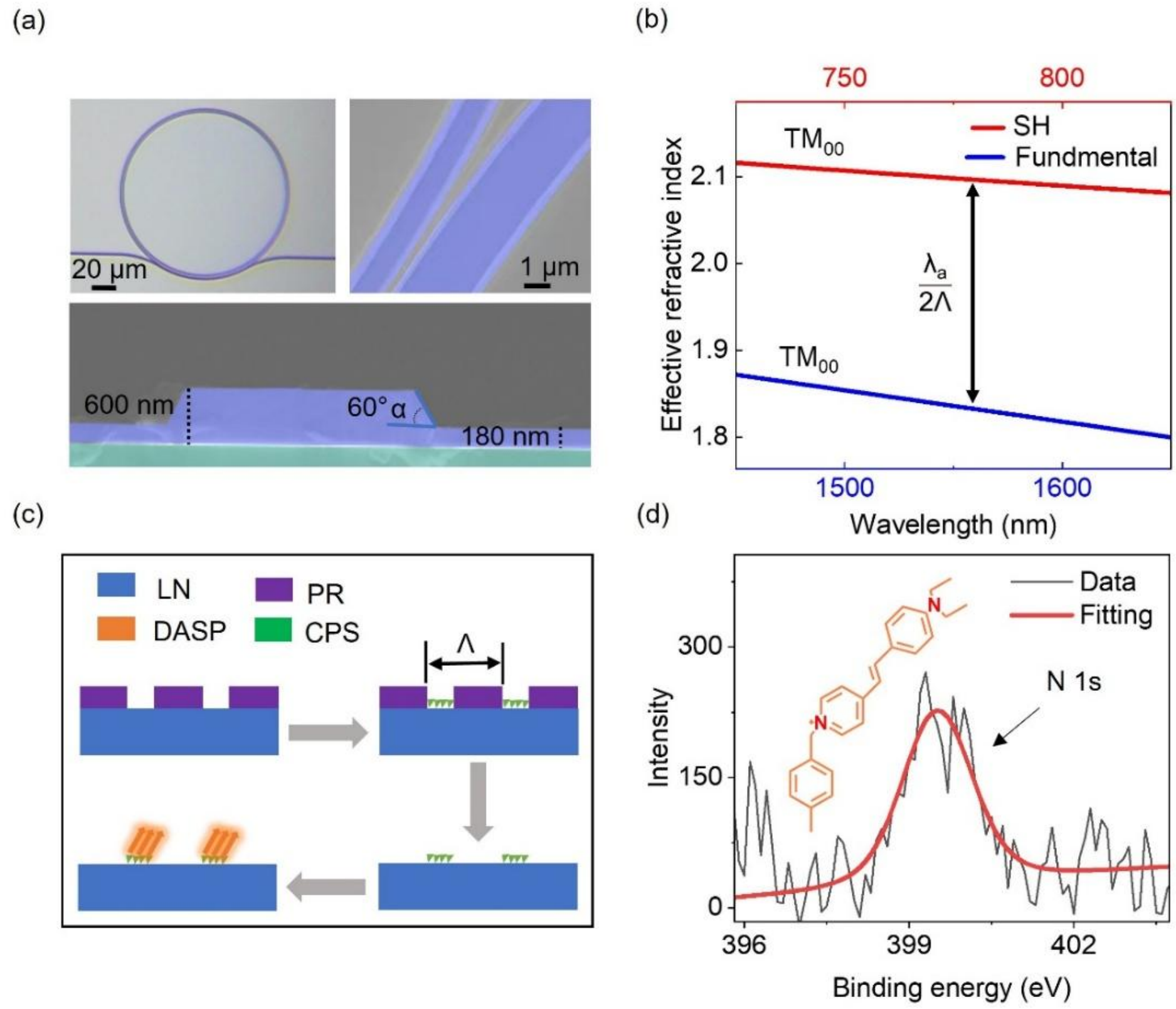

FIGURE 2 | (a) Optical and false-color scanning electron microscopy (SEM) images of the microring resonator. The upper-left panel shows the optical image of the microring resonator; the upper-right and bottom panels are the SEM image of the top-view and the cross-section of the waveguide. (b) Simulated effective refractive indices of the fundamental and SH modes as a function of wavelength. The condition corresponding to a molecular pattern period of Λ = 2.94 μm is indicated by the vertical double-headed arrow. (c) Schematic of the surface molecule patterning process. (d) A high-resolution XPS spectrum of the microring with surface molecular pattern of DASP, showing a binding energy peak at 399.4 eV of nitrogen atoms of DASP. Inset shows the molecular structure.

The fabricated devices were characterized using a continuous-wave (CW) telecom laser (Toptica CTL1550, tunable from 1510 to 1630 nm) and a near-visible laser (Newport TLB6712, tunable from 765 to 781 nm), as shown in Figure 3a. Figures 3b and 3c display the transmission spectra of the pump and SH modes, respectively. The corresponding loaded, intrinsic, and coupling Q-factors were extracted and are displayed in the respective insets. Figure 3d shows the pump spectrum and the corresponding SH spectrum, measured using an optical spectrum analyzer and a spectrograph equipped with an electron-multiplying charge-coupled device (EMCCD), respectively. The SH signal peaks at 775.71 nm under a fundamental pump wavelength of 1551.42 nm. As the power increases, a large number of modes can generate SH. Several modes were studied which generated strong SH signals at low-power pump.

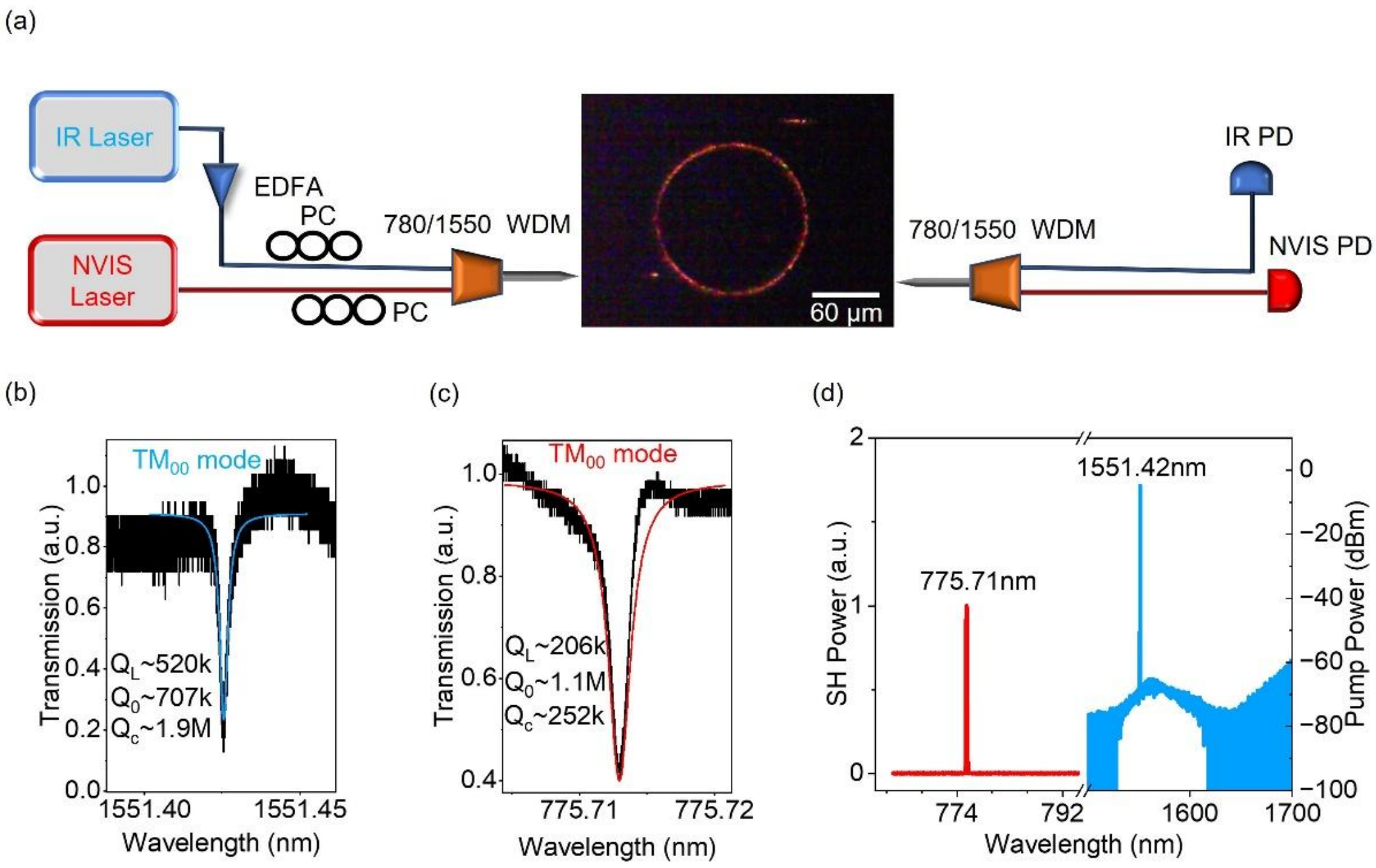

FIGURE 3 | (a) Illustration of experimental setup. EDFA, Erbium-Doped Fiber Amplifier. WDM, wavelength division multiplexer. PD, photodetector. PC, polarization controller. The telecom laser pumped the LN microring resonator, leading to SHG. Inset is a top-viewed optical image of the device with a pump laser at ~1550, showing strong SHG emission. (b-c) Transmission spectra of telecom $TM_{00}$ mode and near-visible $TM_{00}$ mode with extracted loaded, intrinsic, and coupling Q factor. (d) Representative spectrum of the pump and its corresponding second-harmonic signal generated from the devices.

Due to the simultaneous presence of material and geometric dispersion in TFLN microrings, a frequency mismatch inherently exists between the telecom fundamental mode and its corresponding SH mode. The resonance frequency mismatch can be compensated by temperature control of the devices to reach dual resonance condition for efficient SHG (Figure 4a). As plotted in Figure 4b, the SH mode undergoes a faster thermal tuning speed (10.82 pm/°C) than the fundamental mode (2.96 pm/°C). Figure 4c tracks the peak SHG power across varying temperatures under a constant external pump power of 300 μW. As the temperature increases, the SH power first increases and then decreases, reaching the maximum SH output at 29 °C. The temperature dependence of SH output follows a Lorentzian profile with a full width at half-maximum (FWHM) of 2.01 °C. This full evolutionary pathway was dynamically recorded and is presented in Figure S3. Figure 4d shows the measured pump transmission and the corresponding SH response at 29 °C under fine tuning of the pump frequency, confirming the realization of dual resonance of the fundamental and SH modes. Detailed protocols for the data acquisition in Figure 4d are available in the Supporting Information.

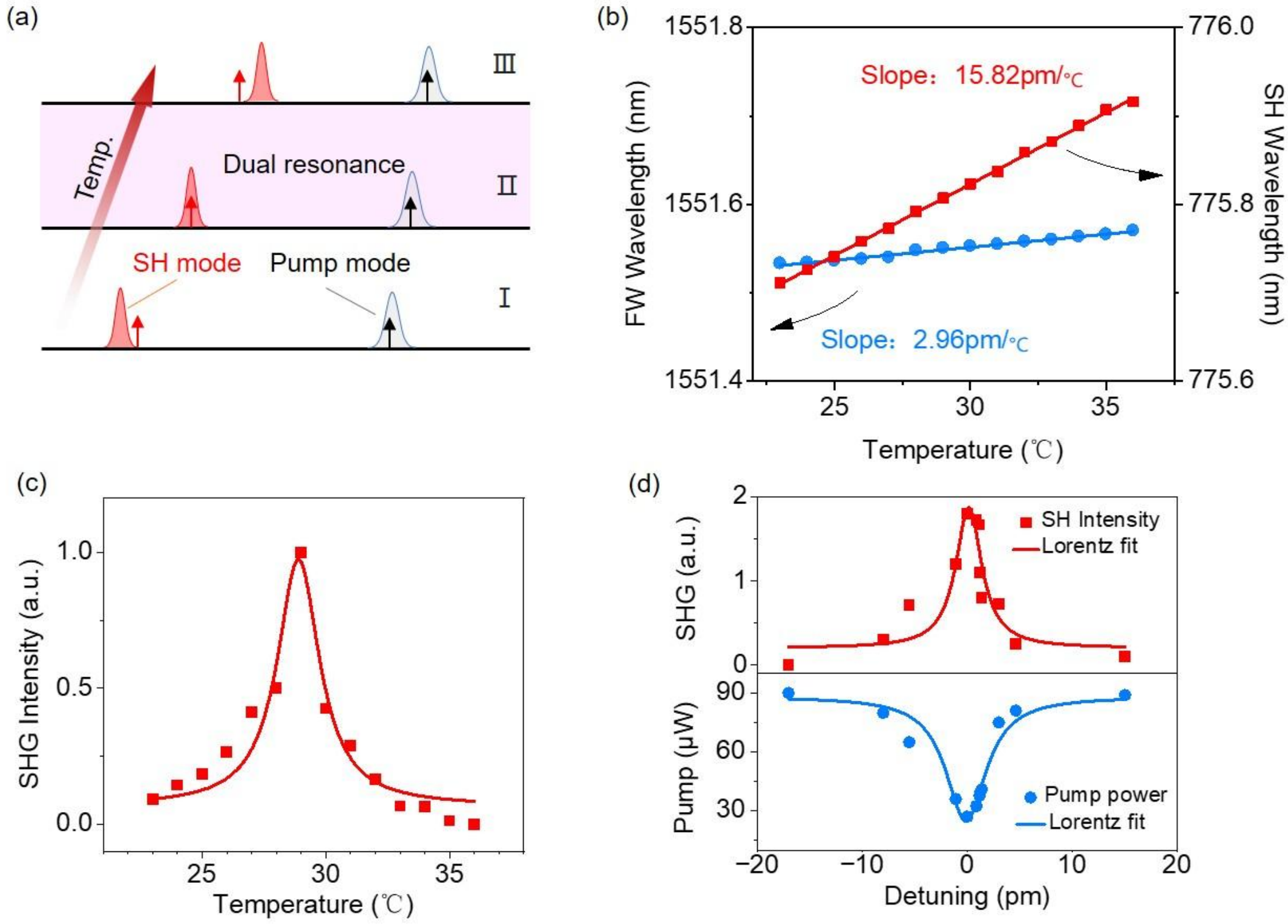

FIGURE 4 | (a) Schematic of dual-resonance phase matching process via temperature control of the device. (b) Resonance shifts versus the temperature for the pump and SH modes, respectively. (c) Max SH power at varying temperatures. A Lorentzian fit is used to extract a FWHM of 2.01 °C. (d) The pump resonance and corresponding SH response at the optimal temperature.

To evaluate the impact of the surface molecular pattern on SHG, we measured the on-chip SH power versus pump power for the same poling-free TFLN device with and without the surface patterned molecules. Figure 5a shows the on-chip SH power as a function of the squared pump power. With the surface patterned molecules, the device starts to generate detectable SH signals at a low pump power of ~45 μW. The normalized conversion efficiency for the device with surface patterned molecules, $\eta = P_{SHG}/P_p^2$, reaches $11.5\%\, W^{-1}$ in the undepleted-pump regime. However, without surface patterned molecules, the same device shows no detectable SH signals in the low pump power range. Only at pump power higher than 0.8 mW can be observed detectable SH signals, with $\eta$ of only $0.029\%\, W^{-1}$ owing to the lack of QPM.

The absolute conversion efficiency (ACE) as a function of the on-chip pump power is shown in Figure 5b. For the device with surface patterned molecules, the SH output begins to saturate at the on-chip pump power exceeds approximately 1 mW. The ACE reaches a maximum value of ~0.016%, showing two-order of magnitude higher than the same device without surface patterned molecules. These results experimentally demonstrate that surface molecular pattern enables QPM for SHG in poling-free TFLN, with about 400 times enhancement in normalized conversion efficiency compared to the same devices without molecular pattern. It is noted that the moderate conversion efficiency is mainly limited by the unoptimized surface density of the molecules which is ~15% of the optimal value (Figure S4) and the consequent enlarged molecular tilt angle. Optimization of the molecular pattern, including the duty circle of the molecular pattern, the surface density of the molecular layer and the molecular hyperpolarizability, would significantly improve the conversion efficiency.

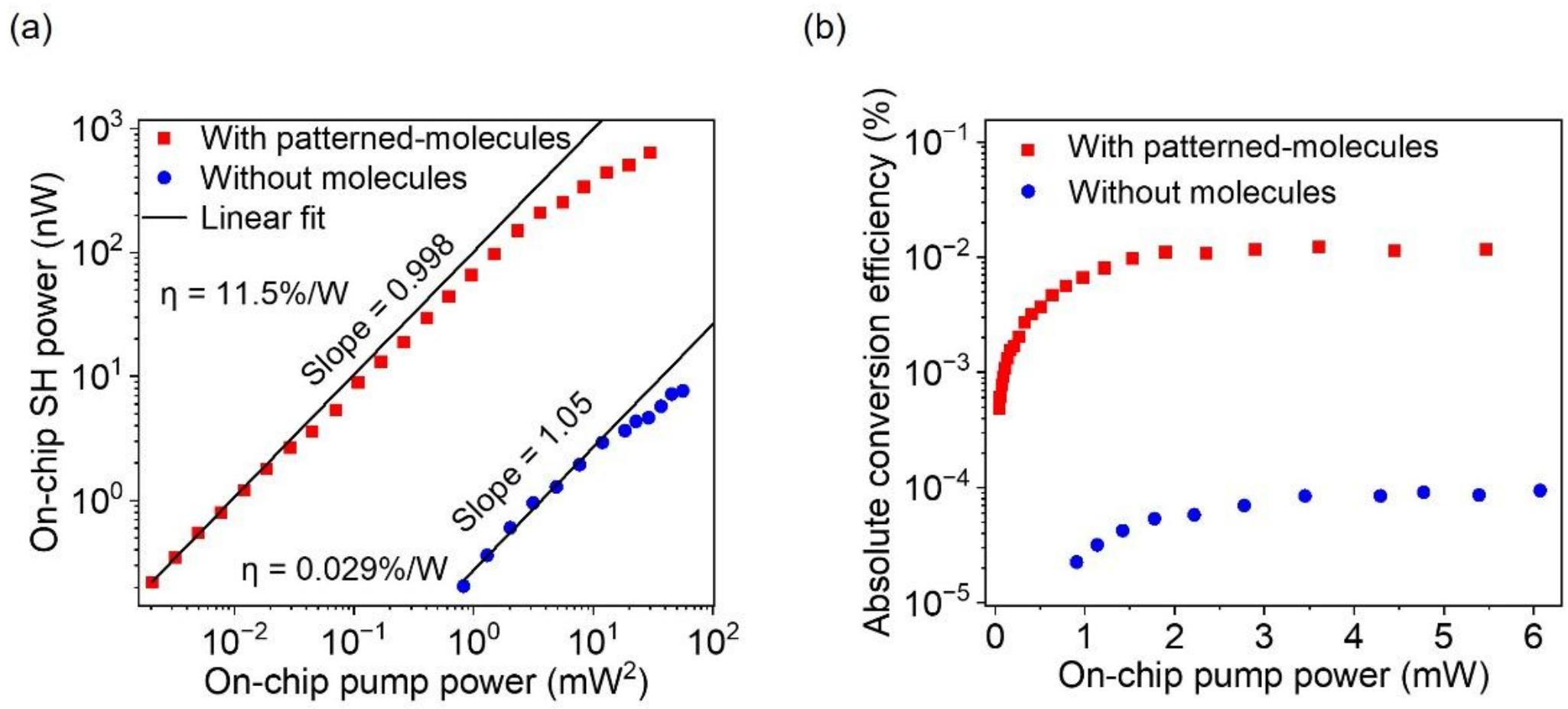

FIGURE 5 | (a) $P_{SHG}/P_p^2$ relation. The normalized conversion efficiencies are 0.029%/W for the device after molecule removal and 11.5%/W for the pattern-surface molecular device. (b) Absolute conversion efficiency (ACE) as a function of pump power.

**4 | Conclusion**

In summary, we have demonstrated a surface molecule patterning strategy to enable QPM in a poling-free TFLN microring for SHG. The surface NLO molecular pattern introduced an interfacial nonlinear grating designed to compensate for the azimuthal mode-number mismatch in the TFLN microring resonator without electric poling. We demonstrated an on-chip normalized conversion efficiency of 11.5% $W^{-1}$ in the device with surface molecular pattern, showing an improvement of ~400 times compared that without surface molecular pattern. Optimization of the molecular pattern, including the duty circle of the molecular pattern, the surface density of the molecular layer and the molecular hyperpolarizability, would further improve the conversion efficiency. This approach offers a versatile route to engineer interfacial nonlinear interactions for nonlinear optics and quantum photonics applications and may be extended to multiple-ring coupled resonators and other platforms.

## 5 | References

## Author contributions

J. Cheng and S. Yu contributed equally to this work. S. Yu fabricated the TFLN mirroring devices. J. Cheng functionalized the devices and conducted the optical measurement. J. Wang conducted the theoretical analysis of QPM. Y. Zhang conducted the simulations. J. Cheng wrote the manuscript with input and discussion from all the authors. The project was under supervision of J. Lu and X. Shen.

## Acknowledgements

This work is supported by the National Natural Science Foundation of China (Grant No. 62275152, 12574362, 62305214 and 92576113), Shanghai Science and Technology Committee (25JD1405800). The facilities used for device fabrication were supported by the ShanghaiTech Material Device Lab (SMDL) and the Soft Matter Nanofabrication Laboratory.

## Data Availability Statement

The data that support the findings of this study are available from the corresponding author upon reasonable request.

## Conflict of interest

The authors declare no conflict of interest.

## Supporting Information

Supporting Information is available from the Wiley Online Library or from the author.

*Supplementary Information for:*

**Quasi-Phase-Matched Second Harmonic Generation in Poling-Free Lithium Niobate Microrings with Surface Molecular Pattern**

## Methods

### Section 1. Device fabrication

The microring resonators were fabricated on commercial NanoLN lithium niobate on insulator (LNOI) wafers, which feature a 600-nm-thick TFLN film on a 2.0-μm buried $SiO_2$ layer over a 525 μm silicon substrate. Microring patterns were defined by electron-beam lithography (EBL) with hydrogen silsesquioxane (HSQ) resist and developed in 25% tetramethylammonium hydroxide (TMAH). An optimized $Ar^+$- based inductively coupled plasma reactive-ion etching (ICP-RIE) process was used to transfer patterns into the TFLN layer. To remove dry-etch-induced redeposition, the chip was cleaned in an $H_2O_2$:$NH_4OH$:$H_2O$ (2:2:1 by volume) solution at 40 °C for 40 min. Remaining HSQ was finally stripped in 6:1 buffered oxide etch (BOE).

### Section 2. Molecular functionalization

To achieve periodic patterning of nonlinear molecules, we first defined patterns on the microring resonator using EBL, an optical image is shown in Fig. S1 (a-b). Nonlinear molecules were functionalized onto the parts of the microring resonators not covered by the photoresist mask. TFLN microring resonators patterned with photoresist were first treated with $O_2$ plasma to generate a high density of hydroxyl groups on the LN surface. The surface was then grafted with a coupling layer using CPS, followed by removal of the photoresist. Next, 60 μL of DASP solution in THF ($1 \times 10^{-4}$ mol/L) was deposited onto the surface of the TFLN chip. The chip was subsequently incubated at 120°C under vacuum for approximately 20 min to accelerate the quaternization reaction. Afterward, the TFLN microring resonators were cooled to room temperature, thoroughly rinsed with tetrahydrofuran, and dried under vacuum at 100°C for 5 min, yielding surface-functionalized TFLN microring resonators.

### Section 3. Q factors characterization

Following surface modification, the resonance wavelength of the LN microring resonators shifted by approximately 0.19 nm, while the Q factor decreased by about one half, as shown in Fig. S2. The observed resonance shift is mainly ascribed to the perturbation of the microring resonators effective refractive index caused by the immobilized molecular layer interacting with the evanescent optical field. In contrast, the reduced quality factor suggests the introduction of additional cavity losses after functionalization, which may originate from enhanced surface scattering due to periodic molecular modification, as well as the absorption loss from the grafted molecules.

**Section 4. The dual-resonance process**

In TFLN, the thermo-optic coefficient of the extraordinary index is significantly larger than that of the ordinary index, and dispersion is stronger at 2ω. As a result, with increasing temperature the SH resonance red-shifts faster than the FW resonance. At a point, the double-resonance condition is then fulfilled, where the SH mode frequency is twice the frequency of the FW mode and both the input FW and generated SH waves are correspondingly resonant. We recorded the dynamic resonance chasing process. The on-chip pump and SH signals were collected with a lensed fiber and sent to a wavelength-division multiplexer (WDM) to separate the two bands. One output was connected to a 1550 nm power meter to measure the on-chip pump power, and the other to an Andor spectrometer to record the relative SH spectral intensity. The laser's automatic wavelength sweep was disabled, and the wavelength was tuned in small increments from the blue- to the red-detuned side while recording simultaneous readings from the power meter and spectrometer. We then raised the chip temperature and repeated the procedure. The recorded data were plotted as scatter points and fitted with Lorentzian line shapes, as shown in Figure S3. As the temperature increases, the FW and SH resonances moved closer to each other and the SH signal strengthens. At a certain temperature, dual resonance was achieved, yielding the maximum SH output. Further heating moves the system pasted this crossing, breaks the dual resonance condition, and the SH signal begins to decrease.

**Section 5. Analysis of the surface molecular density**

The surface density of DASP molecules on TFLN was evaluated via comparison to DASP-functionalized silica surface under the same experimental condition, by XPS. Silica surface possesses a high concentration of surface hydroxyl groups after $O_2$ plasma treatment, yielding a dense, smooth DASP single-molecular layer on the surface after surface functionalization.[1] A similar functionalization process could lead to a lower surface density of DASP molecules on TFLN surface, due to the much less surface hydroxyl groups generated. As shown in Figure S4, DASP-functionalized TFLN surface showed a substantially smaller N 1s peak area about 15% of that of DASP-functionalized silica surface. Further optimization of the surface functionalization would improve the surface molecular density and thus an optimal $\chi_m^{(2)}$, according to $\chi_m^{(2)} \propto N_s \beta_{uuu} \langle cos^3 \theta \rangle$ as mentioned in the main text.

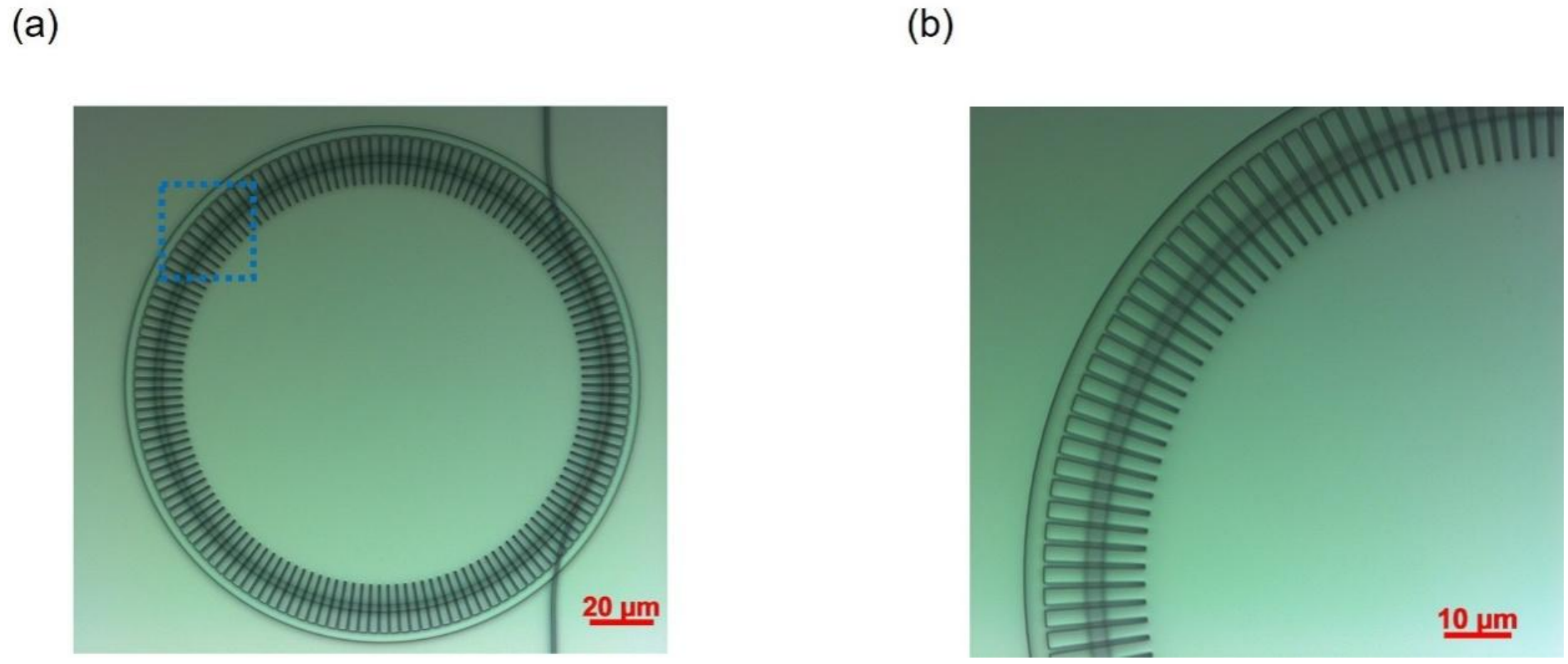

Figure S1. (a-b) optical images of an unpoled TFLN microring resonator patterned with photoresist and its zoomed view.

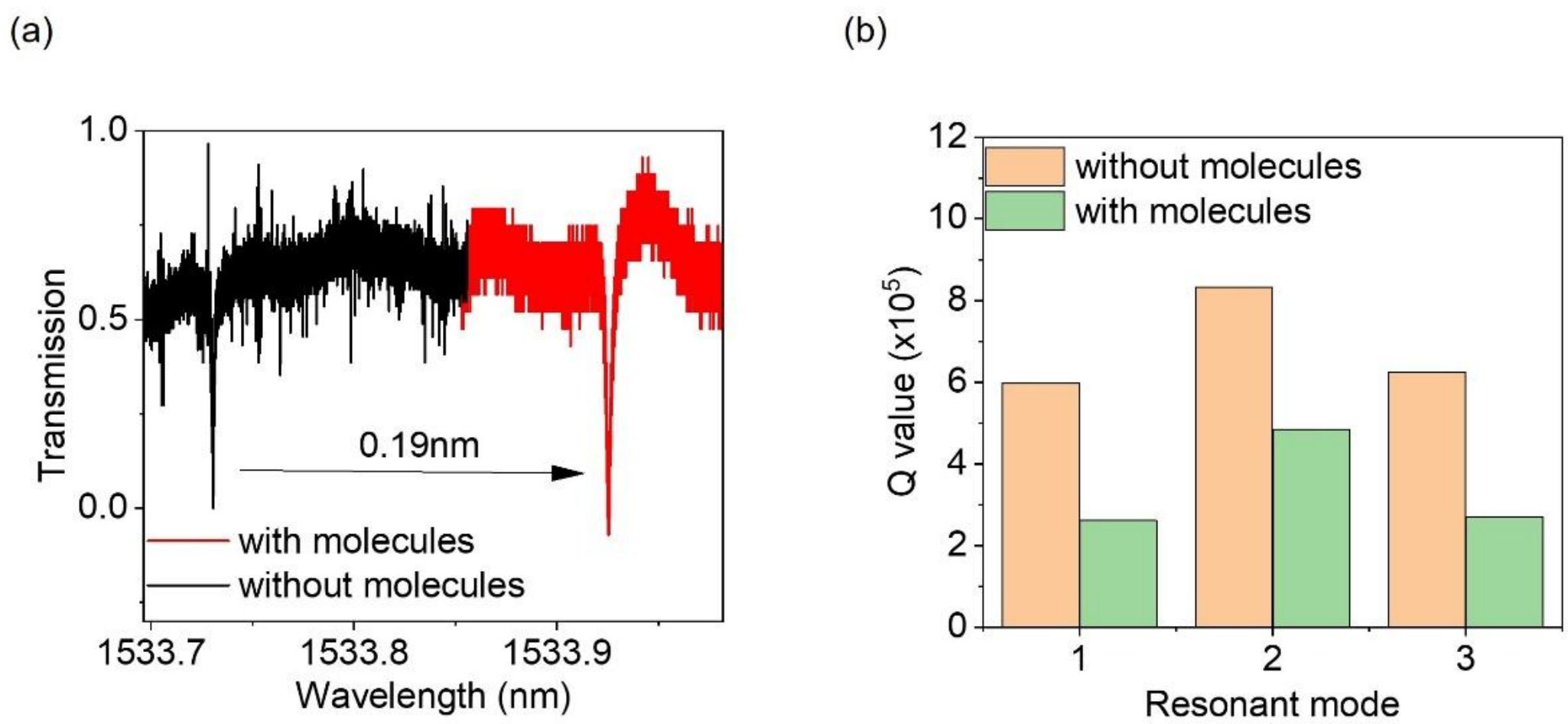

Figure S2. (a) The shift of resonant mode before and after surface functionalization. (b) The measured loaded Q values of three different modes of TFLN microring before and after functionalization.

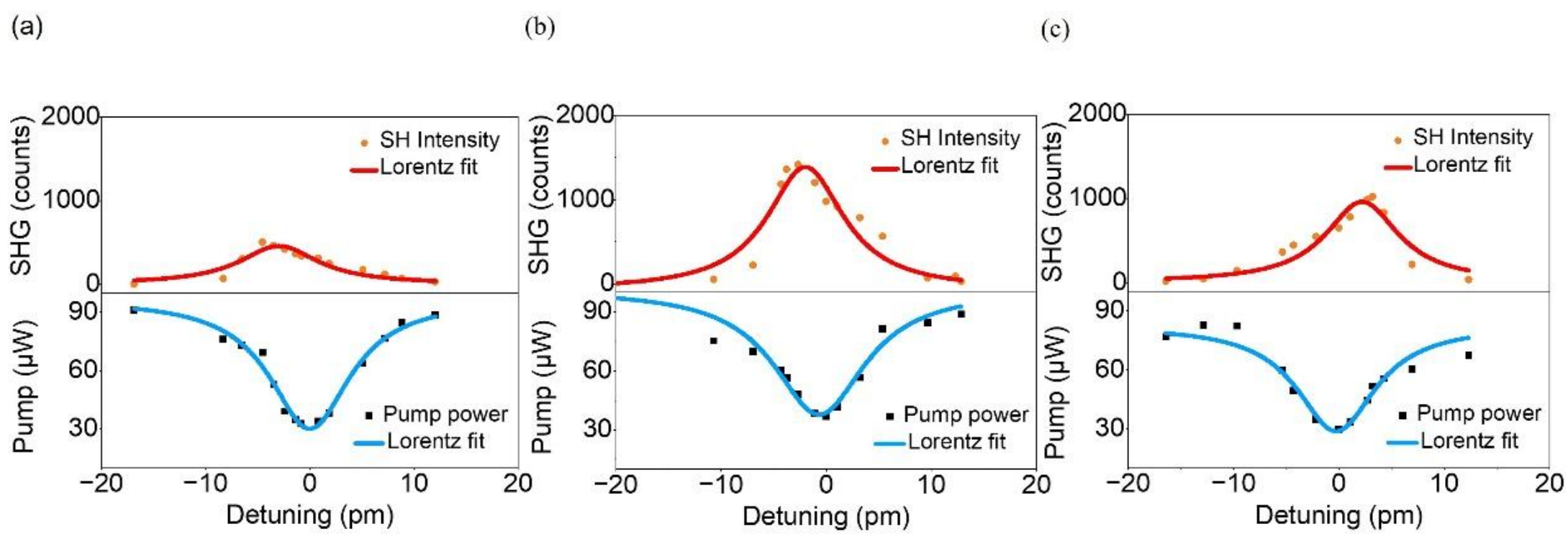

Figure S3. (a-c) Transmission of FW spectra and generated SH signal spectra with different double resonance detuning.

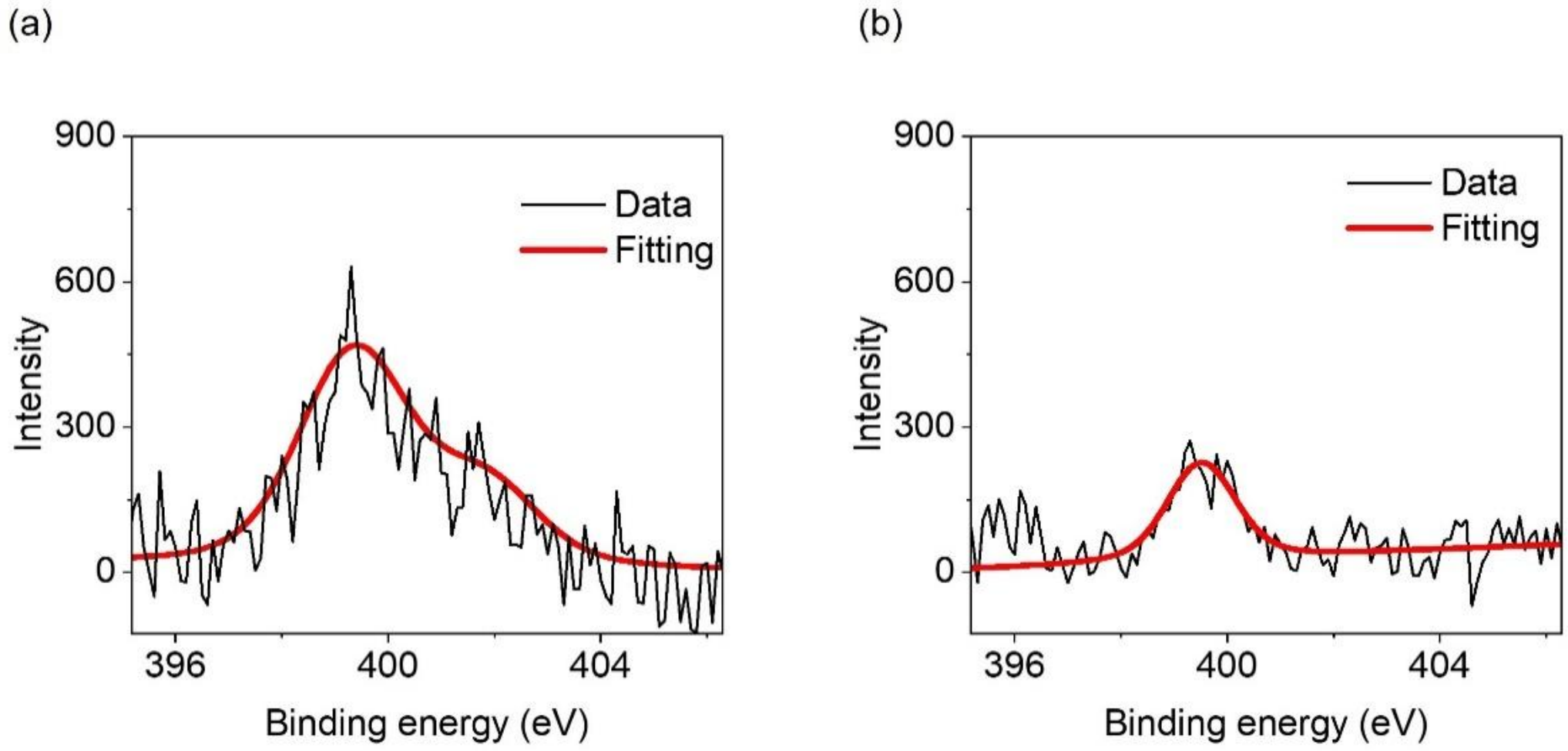

Figure S4. High-resolution N 1s XPS spectra of DASP-functionalized substrates: (a) silica and (b) TFLN. The larger N 1s peak area for the functionalized silica indicates a higher surface molecular density of DASP, while the smaller peak area for the functionalized TFLN.